\documentstyle[12pt]{article}
\begin{document}
\renewcommand{\theequation}{\thesection.\arabic{equation}}
\renewcommand{\section}[1]{\addtocounter{section}{1}
\vspace{5mm} \par \noindent
  {\bf \thesection . #1}\setcounter{subsection}{0}
  \par
   \vspace{2mm} } 
\newcommand{\sectionsub}[1]{\addtocounter{section}{1}
\vspace{5mm} \par \noindent
  {\bf \thesection . #1}\setcounter{subsection}{0}\par}
\renewcommand{\subsection}[1]{\addtocounter{subsection}{1}
\vspace{2.5mm}\par\noindent {\em \thesubsection . #1}\par
 \vspace{0.5mm} }
\renewcommand{\thebibliography}[1]{ {\vspace{5mm}\par \noindent{\bf
References}\par \vspace{2mm}}
\list
 {\arabic{enumi}.}{\settowidth\labelwidth{[#1]}\leftmargin\labelwidth
 \advance\leftmargin\labelsep\addtolength{\topsep}{-4em}
 \usecounter{enumi}}
 \def\newblock{\hskip .11em plus .33em minus .07em}
 \sloppy\clubpenalty4000\widowpenalty4000
 \sfcode`\.=1000\relax \setlength{\itemsep}{-0.4em} }
\newcommand\rf[1]{(\ref{#1})}
\def\nn{\nonumber}
\newcommand{\sect}[1]{\setcounter{equation}{0} \section{#1}}
\renewcommand{\theequation}{\thesection .\arabic{equation}}
\newcommand{\ft}[2]{{\textstyle\frac{#1}{#2}}}

\thispagestyle{empty}

\begin{center}

\vspace{3cm}

{\large\bf `Desert' in Energy or Transverse Space?}\\

\vspace{1.4cm}

{\sc C. Bachas}\footnote{Based on
 talks given at the conferences
`22nd Johns Hopkins workshop' (G\"oteborg, August 1998), 
`Fundamental interactions: from symmetries to
black holes' in honor of Francois Englert (Brussels, March 1999) and
`From Planck Scale to Electroweak Scale' (Bad Honnef, April 1999).}
\\

\vspace{1.3cm}

{\em Laboratoire de Physique Th{\'e}orique
de l' Ecole Normale Sup{\'e}rieure} \\ 
{\em 24  rue Lhomond, 
 75231 Paris Cedex 05, France} \\

\vspace{1.2cm}

\centerline{\bf Abstract}
\vspace{- 4 mm}  \end{center}
\begin{quote}\small
I review  the issue  of string and compactification scales in
the weak-coupling  regimes of string theory. I explain how in the 
Brane World scenario a (effectively) 
 two-dimensional transverse space that is 
hierarchically larger than the string length may  replace
the conventional `energy desert' 
described by  renormalizable supersymmetric QFT. I comment on the
puzzle of unification in this context.
\end{quote}
\vfill

\baselineskip18pt
\addtocounter{section}{1}
\par \noindent
{\bf \thesection . The SQFT Hypothesis}
  \par
   \vspace{2mm} 
\noindent 

 String/M-theory \cite{GSW} is a higher-dimensional theory with a single
dimensionful parameter, which can be taken to be
 the fundamental string tension  or 
the eleven-dimensional Planck scale.
 The theory has on the other hand a large number of `dynamical
parameters'
characterizing  its  many distinct 
semiclassical vacua, such as compactification radii  or  sizes of 
defects localized in the compact space.
 Understanding how the  Standard Model
and Einstein gravity arise at low  energies in one of those vacuum
states  is a central outstanding
problem of String/M-theory.

 The usual hypothesis is that the string, compactification and Planck
scales lie  all close to one another, and that the physics at lower
energies is well described by some  effective  four-dimensional 
renormalizable supersymmetric
quantum field theory (SQFT), which  must include   the Minimal 
Supersymmetric Standard Model (MSSM) and some  hidden sectors. 
I will refer to this picture of the world as the `SQFT hypothesis'.
In this picture the breaking of the residual supersymmetry and the
generation of the electroweak scale are believed to be 
triggered by non-perturbative gaugino 
condensation -- a  story that is however  incomplete because of the 
problems of vacuum stability and of the cosmological constant.

   The minimal version of the SQFT hypothesis  is obtained when there are
no light fields charged under $SU(3)_c\times SU(2)_{ew}\times U(1)_Y$,
besides   those of  the MSSM.
This is  the `energy desert' scenario -- 
a slight misnommer since the `desert'  may  be populated by all sorts of
stuff   coupling   with gravitational strength
to ordinary matter.  The  minimal unification  scenario is supported,
 as is well known \cite{GQW,SU,L},
 by two pieces of strong, though indirect evidence:
 (a) the measured low-energy
gauge couplings  
{\em do meet}  when extrapolated to  higher energies with
the MSSM $\beta$-functions, 
and (b) the energy  $M_{U}\sim 2\times 10^{16} GeV$ at which  they meet 
is in the same  ball-parc  as the string scale. A detailed analysis
within  the weakly-coupled heterotic string \cite{Ka}  leads, in fact,
 to  a  discrepancy of
roughly one order of magnitude between the theoretical point  of 
string unification,  and the one  that  fits  the low-energy data. 
This is a small discrepancy on a logarithmic scale, and it  could be
fixed by small modifications of the minimal
scenario \cite{BFY}.

 Besides being simple and rather natural, 
the minimal SQFT hypothesis makes thus {\it two}
 quantitative predictions which
fit the low-energy data to better than one part in ten.

\vfil\eject


\setcounter{equation}{0}
\section{The Weakly-Coupled Heterotic Theory}

   The SQFT hypothesis is particularly compelling in the context of
the weakly-coupled heterotic string \cite{K}.
Both the graviton and the gauge  bosons
live in this case in the ten-dimensional bulk, and their leading
interactions
are given by the same order in string perturbation theory
(i.e. the sphere diagram). This leads to the universal relation between
the
four-dimensional Planck mass ($M_P$) 
 and the tree-level  Yang-Mills  couplings \cite{G}, 
 \begin{equation}
M_P^2 \sim M_H^2/g^{ 2}_{\rm YM}\ ,
\label{eq:gins}
\end{equation}
independently  of the  details of compactification. 
If we assume that
$g_{\rm YM}\sim o(1)$, then 
 the heterotic string scale ($M_H$) is  necessarily
tied to the Planck scale. Furthermore,
 the standard Kaluza-Klein formula for the
four-dimensional gauge  couplings is
 \begin{equation}
1/g^2_{\rm YM} \sim (R M_H)^6 /g_H^2\ , 
\label{eq:wch}
\end{equation}
with $R$ the typical radius
 of the six-dimensional compact space 
and $g_H$ the
dimensionless string coupling. Pushing 
the Kaluza-Klein scale
($M_{\rm KK} \sim R^{-1}$)  much below $M_H$ requires therefore
a hierarchically-strong  string coupling, and invalidates the
semiclassical
treatment of the vacuum.
 Of course all radii need not be
equal  but,  at least in orbifold
 compactifications,  T-duality allows us
to take them all larger or equal to the  string length,  and then
the above argument forbids any single radius
 from  becoming too large.

  There is actually a  loophole in  the
above reasonning.
If some compact dimensions are much larger
than the heterotic string length, 
loop corrections to the inverse squared gauge couplings
will generically grow like a power of  radius \cite{Ka}.
\footnote{In special models, such as orbifolds without N=2 sectors, 
these  large threshold corrections can be made to vanish
at one-loop. The evolution of gauge couplings with energy is
thus unaffected by  the openning of large extra dimensions \cite{A}.
However, since  $g_H$ must in these models be hierarchically strong, 
the semiclassical string vacuum cannot  be trusted.}
\footnote{ Power corrections
to gauge couplings have been also  recently invoked as a way
to speed up the  unification process \cite{DDG}. } 
It is thus logically conceivable that even though
the observed low-energy gauge
couplings are of order one,  their 
tree-level values  are  hierarchically smaller. 
Since it is the tree-level couplings that enter in the relation
(\ref{eq:gins}),  the heterotic 
string scale could thus in principle be significantly lower than
the four-dimensional Planck mass  \cite{Ba}.

   The main motivation for contemplating such  possibilities 
in the past was  the search for string models
with low-energy  supersymmetry broken spontaneously at  tree level.
Existing heterotic vacua of this type 
 employ a string variant \cite{R} of the
Scherk-Schwarz mechanism \cite{SS},  which breaks supersymmetry in a way
reminiscent of  finite-temperature effects. The scale of (primordial)
breaking is  proportional to an  inverse radius, so that lowering  it to
the electroweak scale requires the openning of extra dimensions at the TeV
--
a feature shown  \cite{ABLT}  to be  generic  in orbifold models.
\footnote{For more general compactifications,
the limit of supersymmetry restoration is also known to be a singular 
limit  \cite{DS}, even though there is no precise relation between
the scale of symmetry breaking and some Kaluza Klein threshold.}

Insisting on  tree-level breaking is, on the other hand,
 only a technical
requirement -- there is no reason why the breaking in nature should not
have a non-perturbative origin. Furthermore,  Scherk-Schwarz
compactification
has not so far lead to any new insights on the problems of vacuum 
selection and stability. Thus, there seems  to be little theoretical
motivation at this point for abandonning the 
SQFT hypothesis, and its successful unification predictions,  
in heterotic string theory.


\section{Brane World and Open String Theory}

   The story is  different in the theory of (unoriented) open and
closed strings, in which gauge and gravitational interactions
have different origins.  While the graviton (a closed-string state) 
lives  in the ten-dimensional bulk, open-string vector bosons 
can  be localized on  defects \cite{DLP} -- the worldvolumes of 
  D(irichlet)-branes \cite{Po}. Furthermore 
while closed strings interact
to leading order via the sphere diagram,  open strings must be attached to
a boundary and thus interact via  the disk diagram which is of higher
order in the genus expansion.
The four-dimensional Planck mass and Yang-Mills couplings
therefore  read 
 \begin{equation}
1/g_{\rm YM}^2 \sim (R_{\parallel}M_I)^{6-n}/g_I \ , \ \ \ 
M_P^2 \sim R_{\perp}^{n}R_{\parallel}^{6-n} M_I^8/g_I^2 ,
\label{eq:typei}
\end{equation}
where $R_\perp$ is the typical radius
 of the n  compact dimensions transverse to the 
brane,   $R_\parallel$ the typical radius
 of the remaining (6-n) compact longitudinal
dimensions, $M_I$ the type-I string scale and $g_I$ the string coupling
constant. As a result  (a) there is no universal relation between
$M_P$, $g_{\rm YM}$ and $M_I$ anymore, and (b) tree-level
gauge couplings corresponding to different sets
of branes have radius-dependent  ratios and need not unify.

   A few remarks before going on. First, we are here discussing a theory
of unoriented strings, because orientifolds \cite{S} 
are required in order
to cancel the tension and  RR charges  of the (non-compact) space-filling
D-branes. Second, using T-dualities we can ensure that 
 both $R_\perp$ and $R_\parallel$ are
greater than or equal to the string scale \cite{DLP}.
 This may take us either to
Ia  or to Ib theory (also called I or I',  respectively) --  I will
not make a distinction between them in what follows.
 Finally, it should be stressed that
D-branes are the only known defects
which  can localize  non-abelian gauge interactions in a perturbative
setting.
 Orbifold fixed points can at most 
`trap'  matter fields and  abelian vector bosons 
(from twisted RR sectors).\footnote{Non-perturbative
 symmetry enhancement is of course
a possibility, as has been discussed for instance in
\cite{AP}. The great success of the perturbative
Standard Model makes one, however,  reluctant to start with  a theory 
 in which $W$ bosons, and all quarks and leptons
do not correspond to  perturbative quanta.}


    Relations (\ref{eq:typei}) tell us that type I string theory is
much more flexible (and  less predictive)  than  heterotic theory.
The string scale $M_I$ is now a free parameter, even if one insists that
both $g_{\rm YM}$ and $g_I$  be  kept fixed and
of $o(1)$.  This added flexibility can be used to remove  the
order-of-magnitude discrepancy
between the unification and string scales \cite{W}. A much more drastic
proposal \cite{ADD,Ly,AADD} is  to
lower $M_I$  down to the  experimentally-allowed limit $\sim
o({\rm TeV})$. Keeping for instance $g_I$,
$g_{\rm YM}$ and $R_\parallel M_I$ of order one, leads to the condition
 \begin{equation}
R_\perp^n \sim M_P^2/M_I^{2+n} .
\label{eq:mm}
\end{equation}
A TeV string scale would then require from  n=2 millimetric 
to  n=6 fermi-size dimensions transverse to our Brane World --  the
relative weakness of gravity being in this picture 
attributed to  the transverse spreading of
gravitational flux.

  What has brought this idea \footnote{For early discussions
 of a Brane Universe see \cite{RS}.}
 into sharp focus \cite{ADD} was 
(a) the realization 
 that submillimeter   dimensions are not at present 
ruled out by mesoscopic gravity
 experiments,\footnote{That such experiments do not rule out
light scalar particles, such as axions,  
with gravitational-force  couplings
and Compton wavelengths of  a millimeter or less, had been already
appreciated in the past \cite{MW}.  The Kaluza-Klein excitations of the
graviton are basically subject  to the same  bound.}
 and (b) the hope  that 
lowering $M_I$ to the TeV scale may lead to a new understanding   of
the gauge hierarchy. Needless to say that  a  host of
constraints (astrophysical and cosmological bounds, proton
 decay, fermion masses etc.) will make 
realistic model building a very strenuous exercise indeed. 
Finding type I vacua with three chiral families of quarks and leptons
is already a
 non-trivial problem by itself \cite{KT}.   None of
these difficulties seems,  however,  {\it a priori} fatal
to the Brane World idea,  even in its
 most extreme realization \cite{checks}.

\setcounter{equation}{0}
\section{Renormalization Group or Classical Supergravity?}

   Although the type I string scale  could lie anywhere below the
 four-dimensional Planck mass,\footnote{Arguments in favour of an
 intermediate string scale were given  in \cite{BIQ}.} 
 I will now focus on the extreme case
 where it is close to its experimental lower limit,
 $M_I \sim o({\rm TeV})$. Besides being a natural starting point for
 discussing the question of the gauge hierarchy, this has also the
 pragmatic  advantage of bringing string physics
 within the reach of future acceleretor experiments.
 This extreme choice is at first sight
antipodal to the minimal SQFT hypothesis~:
 the MSSM is a stable 
renormalizable field theory, and yet one  proposes  to shrink 
its range of validity  to  one order of magnitude at most!
 Nevertheless, as I will now argue, the Brane World and SQFT
 scenaria
 share many common features   when the number of large transverse
 dimensions in the former  is exactly  two \cite{B,AB}.

   The key feature of the SQFT hypothesis is that low-energy
   parameters receive large logarithmic corrections, which are
  effectively resummed  by the equations of the Renormalization Group. 
  This running with energy can  account for  the observed values of the
   three gauge  couplings,  and of the
   mass matrices of quarks and leptons, in a way that is relatively
   `robust'.\footnote{One  must 
of course assume initial conditions for the RG equations, typically
 imposed by unification and by  discrete symmetries, but there is  no
need
   to know in greater detail  the physics in the ultraviolet regime.}
 Furthermore the logarithmic sensitivity of parameters
 generates naturally  hierarchies of
 scales, and has been  the key ingredient in all
  efforts to understand the origin of 
 the $M_Z/M_P$  hierarchy in the past \cite{N}.

 Consider now  the Brane World scenario. The parameters of the
 effective Brane Lagrangian are dynamical open- and closed-string
 moduli. These latter, denoted collectively by $m_K$, 
 include the dilaton,
 twisted-sector massless scalars, the metric of the transverse space etc.
Their vacuum expectation values are constant  along the four non-compact
 space-time dimensions, but  vary  generically as a function of the
 transverse coordinates $\xi$.
For weak  type-I string coupling and large transverse space
these   variations can be
 described by a Lagrangian of the (schematic) form
 \begin{equation}
{\cal L}_{\rm bulk} + {\cal L}_{\rm source}
\sim \int d^n\xi \; \Bigl[ {1\over g_I^2}
(\partial_\xi  m_K)^2 +  {1\over g_I} \sum_s f_s(m_K) 
\delta(\xi-\xi_s)\Bigr]
.
\label{eq:bb}
\end{equation}
This  is a  supergravity Lagrangian reduced to the
n large transverse dimensions, and  
coupling  to D-branes and orientifolds which act as  sources localized
at transverse  positions 
 $\xi_s$.\footnote{In the general case there could be
also branes extending only partially into the large 
transverse bulk. Our discussion can be
adapted easily to take those into account.}
 The couplings
$f_s(m_K)$ may vary  from source to source -- they can for instance
depend on open-string moduli --  and are subject to global
consistency  conditions. What is  important, however, to us
 is that they are
{\it weak} in the type-I limit, leading to weak  variations, 
 \begin{equation}
m_K(\xi) = m_K^0 + g_I\; m_K^1(\xi) + \cdots  , 
\label{eq:bbb}
\end{equation} 
with  $m_K^0$  a constant, $m_K^1$  a sum of Green's functions etc.
For $n=2$ dimensions the leading variation
$m_K^1$  grows  logarithmically with the size  
 of  the transverse space, $R_\perp$. Since our
Standard Model parameters will be  a function of the moduli evaluated at
the position of our Brane World, they will have  logarithmic
sensitivity on $M_P$ in this case, very much like  the (relevant)
parameters of  a  supersymmetric renormalizable QFT. Similar
sensitivity will occur even if $n>2$,  as long as  some twisted
moduli propagate  in only two extra large dimensions.

   Let me now discuss the validity of the approximation (\ref{eq:bb}).
The bulk  supergravity Lagrangian  receives 
both $\alpha^\prime$ and higher-genus corrections, but these involve 
higher derivatives of fields and should be 
negligible for moduli  varying logarithmically over distance scales
$\gg \sqrt{\alpha^\prime}$.
 The source functions, $f_s(m_K)$, are also
in general  modified by such corrections -- our $\delta$-function
approximation is indeed only valid to within $\delta\xi\sim
o(\sqrt{\alpha^\prime})$. Such  source modifications can, however, be
absorbed
into boundary conditions for the classical field equations  
at the special marked points $\xi_s$. The situation thus looks (at
 least superficially) analogous to that prevailing under 
 the SQFT hypothesis~: large  corrections to low-energy parameters
 can be in both cases  resummed by  differential equations
 with  appropriate boundary conditions. There are, to be sure,
 also important  differences~: in particular,
 the  Renormalization Group
 equations are  first order  differential equations
 in  a single (energy) scale
 parameter, while the classical  supergravity equations are 
 second-order and depend on the two coordinates of the large
 transverse  space.

The   analogy between energy and
transverse distance is also reminiscent of the
holographic idea \cite{hol}, considered  in the context of
 compactification in \cite{RSV}. It is, however, important to stress
that our discussion here stayed
pertubative (and there was no large-N limit involved).
 I have just tried to argue  that 
large string-loop corrections to the  parameters of a 
 brane action can, in appropriate settings,
be calculated reliably
 as the sum of two superficially similar effects:
 (a)   RG running from some low energy scale  up to  string scale, and 
 (b)  bulk-moduli  variations over a transverse two-dimensional
space of size much  greater
 than string length. The two corresponding
regimes -- of renormalizable QFT and of reduced
classical supergravity --
 are a priori different  and need not  overlap.

\section{The Puzzle of Unification}

  The logarithmic sensitivity  of brane parameters on
$R_\perp$ can be used to generate  scale hierarchies dynamically,
exactly as with renormalizable QFT.  Gauge dynamics   on
a  given brane, for example, can  become strong as the transverse
space  expands to a hierarchically large size, thereby inducing  gaugino
condensation and possibly supersymmetry breaking. Rather than discussing
such
scenaria further, I would now like to return to the main piece of 
 evidence in favour
of the SQFT hypothesis~: the apparent unification of the Standard
Model  gauge couplings.
Can their  observed low-energy values
 be understood \cite{B,ABD}  in an equally robust and
controlled manner, as coming from logarithmic variations in the
(real)  space transverse
 to our Brane World~?\footnote{For another recent
idea see \cite{I}.} 
I dont yet  know the answer to this important question, but let me at
least refute the following possible objection~:
since the three gauge  groups of the
Standard Model  live at the same point in transverse space
(or else  matter  charged under two of them would have been
ultraheavy)  how can real-space variations split their coupling constants
apart~? This objection would have been, indeed, fatal  if all gauge
couplings 
were determined by the same combination of bulk fields. 
This is fortunately not the case~: scalar moduli from twisted sectors
of  orbifolds have been, for instance, shown to have non-universal
couplings to gauge fields living on the same brane \cite{AFIV,ABD}. 
The logarithmic variations of such 
fields could  split the three Standard Model
gauge couplings apart,  although it is unclear
why this splitting should be in  the right proportion.

\vspace{.75cm}
\vfil\eject


\noindent {\bf Acknowledgements}:
I thank the organizers of the G\"oteborg, Brussels and Bad Honnef meetings
for the invitations to speak, and in particular Fran\c cois Englert
for teaching us all that `physics is great fun'.
I also thank G. Aldazabal, C. Angelantonj, A. Dabholkar, 
M. Douglas, G. Ferretti, B. Pioline,  A. Sen and H. Verlinde
for discussions, and the
ICTP in Trieste for hospitality while this talk was being written up. 
Research  partially supported by  EEC grant
TMR-ERBFMRXCT96-0090.



\end{document}